\documentclass[conference]{IEEEtran}
\IEEEoverridecommandlockouts
\usepackage{cite}
\usepackage{amsmath,amssymb,amsfonts}
\usepackage{algorithmic}
\usepackage{graphicx}
\usepackage{textcomp}
\usepackage{xcolor}
\usepackage{bm}
\usepackage{booktabs}
\usepackage{tabularx}
\usepackage{xspace}
\usepackage{tabularx}
\usepackage{booktabs}
\usepackage{ragged2e}   
\usepackage{array}      
\def\BibTeX{{\rm B\kern-.05em{\sc i\kern-.025em b}\kern-.08em
    T\kern-.1667em\lower.7ex\hbox{E}\kern-.125emX}}

\begin{document}

\title{Learnable Multi-level Discrete Wavelet Transforms for 3D Gaussian Splatting Frequency Modulation\\
}

\author{
\IEEEauthorblockN{
Hung Nguyen,
An Le,
Truong Nguyen
}
\IEEEauthorblockA{
\textit{Video Processing Lab, UC San Diego} \\
\{hun004, d0le, tqn001\}@ucsd.edu
}
}

\maketitle

\begin{abstract}
3D Gaussian Splatting (3DGS) has emerged as a powerful approach for novel view synthesis. However, the number of Gaussian primitives often grows substantially during training as finer scene details are reconstructed, leading to increased memory and storage costs. Recent coarse-to-fine strategies regulate Gaussian growth by modulating the frequency content of the ground-truth images. In particular, AutoOpti3DGS employs the learnable Discrete Wavelet Transform (DWT) to enable data-adaptive frequency modulation. Nevertheless, its modulation depth is limited by the 1-level DWT, and jointly optimizing wavelet regularization with 3D reconstruction introduces gradient competition that promotes excessive Gaussian densification. In this paper, we propose a multi-level DWT-based frequency modulation framework for 3DGS. By recursively decomposing the low-frequency subband, we construct a deeper curriculum that provides progressively coarser supervision during early training, consistently reducing Gaussian counts. Furthermore, we show that the modulation can be performed using only a single scaling parameter, rather than learning the full 2-tap high-pass filter. Experimental results on standard benchmarks demonstrate that our method further reduces Gaussian counts while maintaining competitive rendering quality.
\end{abstract}


\section{Introduction}

3DGS \cite{3DGS} has emerged as a dominant method for 3D scene reconstruction from multi-view 2D images, enabling high-quality novel view synthesis with efficient training. Owing to its strong rendering fidelity and computational advantages, it has been adopted across a wide array of domains \cite{DynaGSSLam, SplatSDF, MonoSelfRecon}.

By representing a scene as a collection of anisotropic Gaussian primitives, 3DGS significantly accelerates optimization and improves visual quality compared to its predecessor, NeRF \cite{NeRF}. Nevertheless, the number of Gaussians is not directly constrained during training and often increases substantially as the model captures finer scene details. Controlling this growth is therefore critical to reducing GPU memory consumption and storage requirements, particularly for deployment on resource-limited or edge platforms \cite{Compressed-3DGS, Reduced-3DGS}. A more compact Gaussian representation also benefits downstream tasks that depend on per-Gaussian embeddings \cite{LangSplat, LEGaussians, WildGaussians, E-D3DGS}.

Recently, multiple efforts have been aimed at optimizing Gaussian counts. In this work, we focus on coarse-to-fine strategies, as they are prior-free and explicitly regulate structural growth by modulating the frequency content of the ground-truth images. By delaying the introduction of fine details, such methods reduce premature over-densification in homogeneous regions, reconstructing high-frequency (HF) structures only when introduced, thereby reducing unnecessary Gaussian proliferation. Opti3DGS \cite{Opti3DGS} proposes blurring the input images with progressively smaller filter sizes as training progresses. However, this approach relies on manually predefined blur types and schedules, which are not dataset-adaptive and may limit rendering quality. AutoOpti3DGS~\cite{AutoOpti3DGS} instead performs frequency modulation using the Discrete Wavelet Transform (DWT). The high-pass analysis filter is initialized to zero, producing an initially low-frequency reconstruction from the Inverse Transform. The reconstruction is gradually restored through a wavelet regularization objective that encourages approximate Perfect Reconstruction~\cite{wavelet-book}. This enables data-adaptive coarse-to-fine modulation and reduces Gaussian counts while maintaining rendering fidelity.
Nevertheless, the method is restricted to a 1-level DWT, limiting the depth of the modulation curriculum since substantial HF content remains in the ground-truth images. Moreover, the wavelet regularization is optimized jointly with the 3DGS reconstruction loss in different domains, potentially introducing competing gradients that might trigger excessive Gaussian densification.

In this paper, we build upon AutoOpti3DGS with two key extensions. Firstly, we generalize the 1-level DWT to multi-level, enabling a deeper frequency modulation curriculum. By recursively decomposing the LL subband, the initial training images become progressively coarser, leading to stronger early-stage low-frequency emphasis. This consistently results in further reductions in Gaussian counts. Secondly, to mitigate conflicting gradients between the wavelet regularization and the reconstruction objective, we reduce the number of learnable wavelet parameters. Specifically, for the 2-tap case, we show that learning only a scaling parameter for the high-pass filter is sufficient. This removes one degree of freedom from the optimization, alleviating gradient competition across domains. As a result, the model can focus more on stable reconstruction while preserving the desired frequency modulation effect. In summary, the contributions are as follows:

\begin{itemize}
    \item We extend AutoOpti3DGS from 1-level to multi-level DWT, constructing a deeper frequency modulation curriculum that produces coarser initial supervision and consistently reduces Gaussian counts.
    \item We show that the modulation can be performed using only a single scaling parameter, rather than learning the full 2-tap high-pass filter.
    \item The proposed method demonstrates to further reduce Gaussian counts, while maintaining rendering quality across benchmark datasets.
\end{itemize}

\section{Related Works}

\textbf{Discrete Wavelet Transform (DWT) for Differentiable Novel View Synthesis (NVS).}
Among recent differentiable NVS approaches, NeRF~\cite{NeRF} and 3DGS \cite{3DGS} have emerged as two dominant paradigms. 
Within NeRF-based frameworks, Rho et~al.~\cite{MaskedWavelet} apply wavelet-based masking to learn a compact representation. 
WaveNeRF~\cite{WaveNeRF} and TriNeRFLet~\cite{TriNeRFLet} incorporate the multi-level DWT to enhance high-frequency (HF) detail preservation. 
DWTNeRF~\cite{DWTNeRF} performs wavelet-domain supervision to reduce HF overfitting in sparse-view settings~\cite{INGP}. Within 3DGS-based frameworks, MW-GS~\cite{MicroMacro} and Wavelet-GS \cite{Wavelet-GS} applies the DWT to simultaneously model coarse and fine details. DWTGS \cite{DWTGS} performs low-frequency-centric wavelet-domain supervision to suppress HF hallucinations under sparse views. Conversely, under sufficient views, 3D-GSW \cite{3D-GSW} utilizes HF-centric wavelet loss to improve detail fidelity. WaveletGaussian \cite{WaveletGaussian} performs wavelet-domain diffusion to efficiently generate pseudo samples for 3DGS-based object reconstruction. AutoOpti3DGS \cite{AutoOpti3DGS} applies the learnable DWT for coarse-to-fine frequency modulation of 3DGS input images, achieving Gaussian count optimization while maintaining reasonable rendering quality. In this work, we extend it with the multi-level DWT for a prolonged modulation curriculum, as well as additional analyses on filter choices.

\textbf{Learnable DWT.} Multiple efforts have aimed at learning the optimal, data-adaptive DWT filters. MLWNet~\cite{MLWNet} employs the learnable DWT to learn task-adaptive representations for motion deblurring. Subsequent works replace CNN pooling layers with the learnable DWT to achieve lossless and adaptive feature extraction. uWu~\cite{UwU} optimizes the wavelets with a relaxed Perfect Reconstruction condition~\cite{wavelet-book} objective, after initializing from orthogonal versions. Later extensions~\cite{Orthogonal-LatticeUwU, Biorthogonal, BiorthogonalEUSIPCO, Stopband, Wavelet3DSeg, WaveletEye} adopt orthogonal lattice and biorthogonal wavelet structures, as well as lifting schemes, to reduce filter complexity or enhance design flexibility. Generally, these methods focus on representation learning. Differently, our work and the predecessor AutoOpti3DGS \cite{AutoOpti3DGS} employ the learnable DWT as a differentiable image modulator, encouraging 3DGS to learn an efficient 3D scene representation.

\section{Preliminary Background}

\subsection{3D Gaussian Splatting} \label{sec_3dgs}

From a set of multi-view 2D images of a scene, 3DGS~\cite{3DGS} reconstructs the scene in 3D by modeling it as a collection of anisotropic 3D Gaussians. Once the scene representation is learned, novel views can be synthesized by rendering from arbitrary viewpoints. Each Gaussian is characterized by learnable parameters: center $\bm{\mu}$, opacity $\sigma$, covariance matrix $\boldsymbol{\Sigma}$, and color $\mathbf{c}$. 

To optimize these parameters, the following differentiable objective is employed:
\begin{equation} \label{eq_3dgs_loss}
    \mathcal{L}_{\text{3DGS}} = (1-\lambda)\mathcal{L}_1(\mathbf{X}^{\text{gt}}, \mathbf{X}) + \lambda \mathcal{L}_{\text{{D\_SSIM}}}(\mathbf{X}^{\text{gt}}, \mathbf{X})
\end{equation}
where $\mathbf{X}^{\text{gt}}$ and $\mathbf{X}$ denote the ground-truth and rendered images from the same camera viewpoint, respectively. $\mathcal{L}_1$ corresponds to the pixel-wise mean absolute error, while $\mathcal{L}_{\text{D\_SSIM}}$ captures perceptual similarity. The parameter $\lambda$ controls the trade-off between the two loss terms.

\subsection{Discrete Wavelet Transforms (DWT)} \label{sec_dwt}

\begin{figure}[thbp]
    \centering
    \includegraphics[width=0.90\linewidth]{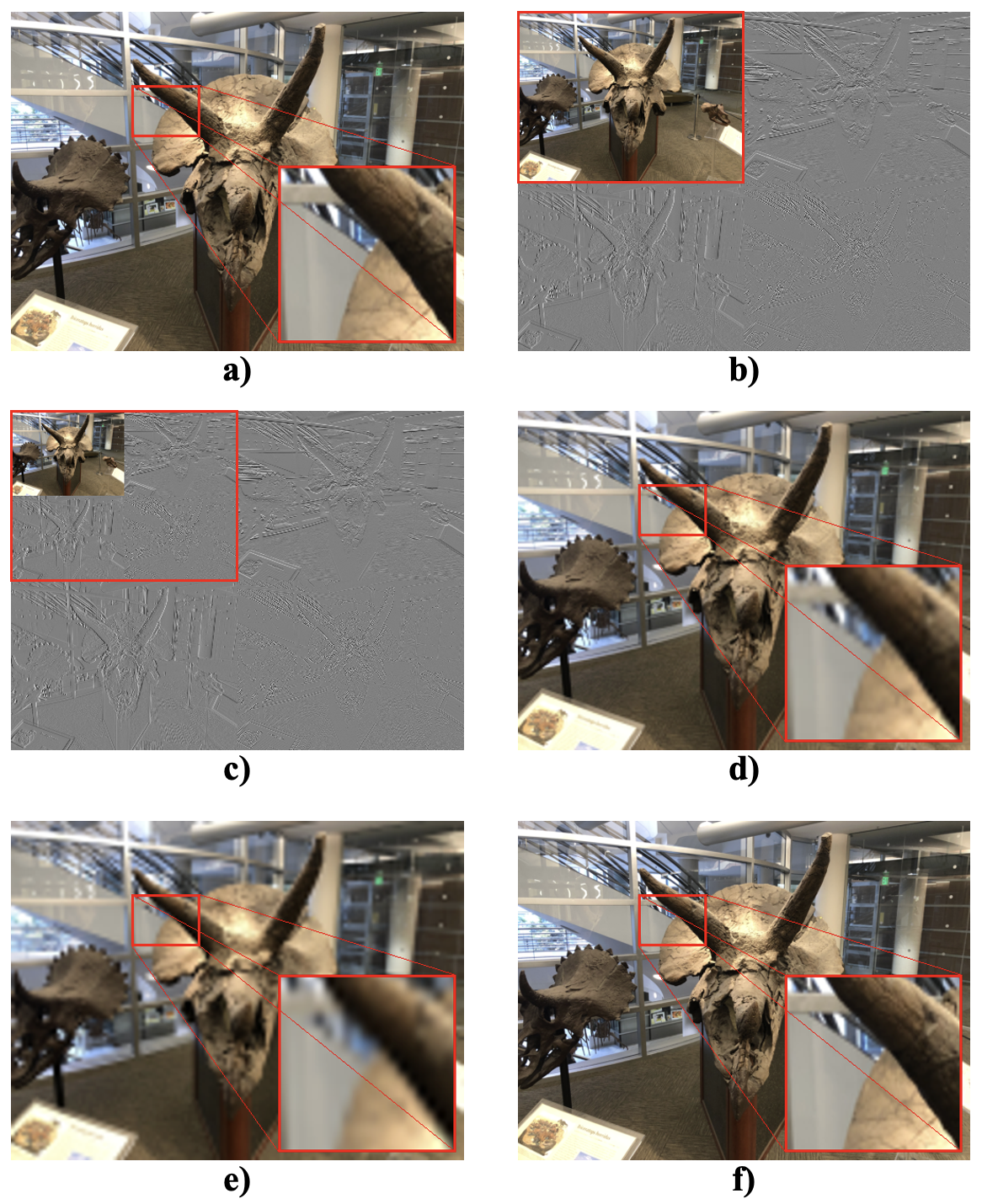}
    \caption{Illustration of the 2D DWT operations. (a) Original image. (b) 1-level DWT subbands. (c) 2-level DWT obtained by further decomposing the 1-level LL subband. (d, e) Enlarged 1- and 2-level LL subbands, respectively. The higher-level LL subbands are coarser. (f) Reconstructed image using PR-satisfying wavelets.}
    \label{fig_vis_dwt}
\end{figure}

\begin{figure*}[htbp]
    \centering
    \includegraphics[width=\linewidth]{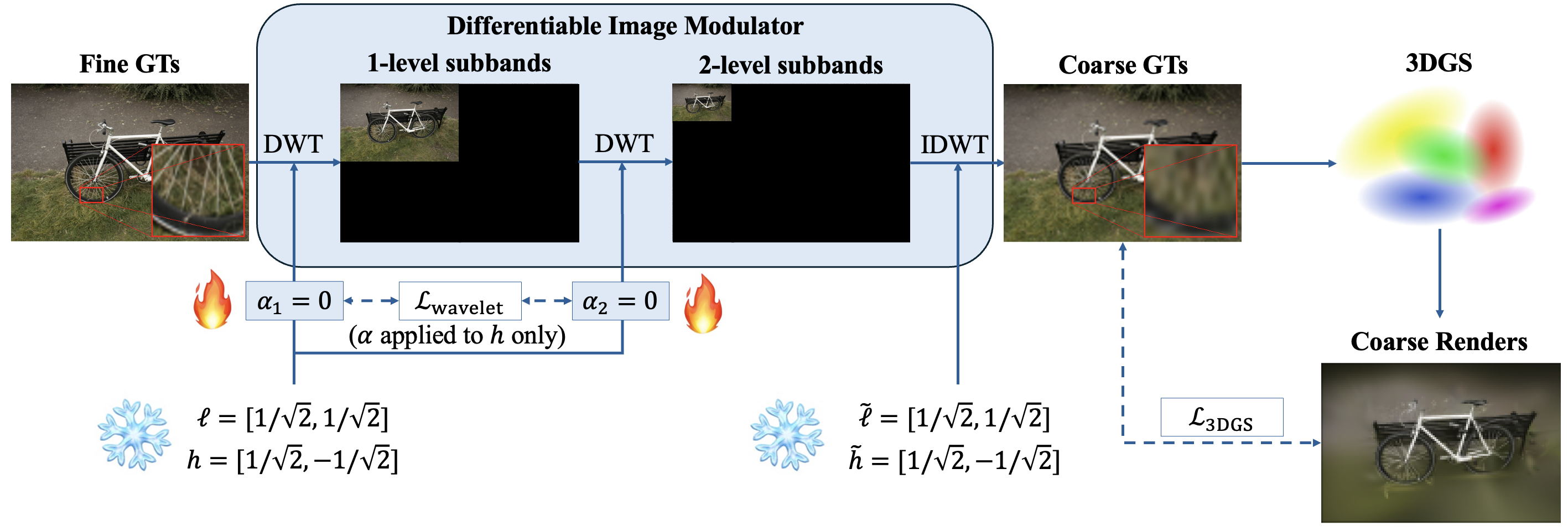}
    \caption{Overview of our framework. The multi-level DWT is employed as a differentiable image modulator. We freeze the original Haar filters and introduce a scaling parameter $\alpha$ on the high-pass analysis filters. When $\alpha=0$, all HF subbands vanish, yielding a coarse IDWT reconstruction, to be used as ground-truths for early-stage 3DGS. A PR-enforcing loss regularizes $\alpha$, progressively restoring high frequencies for automatic coarse-to-fine modulation.}
    \label{fig_framework}
\end{figure*}

The 1D Forward DWT applies a pair of analysis filters, a low-pass filter $\ell$ and a high-pass filter $h$, to a 1D signal $\mathbf{x}$, resulting in the approximation coefficient $\mathbf{x}_L$ and detail coefficient $\mathbf{x}_H$:

\begin{equation}
\mathbf{x}_\text{L} = (\mathbf{x} * \ell)\downarrow_2, \quad
\mathbf{x}_\text{H} = (\mathbf{x} * h)\downarrow_2
\end{equation}
where $*$ denotes the convolution operator and $\downarrow_2$ denotes downsampling by a factor of 2.

Given the coefficients, the 1D Inverse DWT provides the reconstructed signal $\hat{\mathbf{x}}$:
\begin{equation}
\hat{\mathbf{x}}
= \bigl(\uparrow_2\!\mathbf{x}_\text{L}\bigr) * \tilde{\ell}
\;+\;
\bigl(\uparrow_2\!\mathbf{x}_\text{H}\bigr) * \tilde{h}.
\end{equation}
where $\uparrow_2$ denotes upsampling by 2, and $\tilde{\ell}$ and $\tilde{h}$ are the low-pass and high-pass synthesis filters, respectively.

When the analysis and synthesis filters satisfy the Perfect Reconstruction (PR) conditions~\cite{wavelet-book}, the original signal is exactly recovered, i.e., $\hat{\mathbf{x}} = \mathbf{x}$ holds. Figure \ref{fig_framework} shows the coefficients of the Haar, the simplest PR-satisfying wavelet. Further details on the PR property are provided in Section \ref{sec_method}.



Given a 2D image $\mathbf{X}$, the 2D extension of the 1D Forward DWT yields four analysis filters. These filters can be denoted as $\mathbf{K}_{\text{LL}}$, $\mathbf{K}_{\text{LH}}$, $\mathbf{K}_{\text{HL}}$, and $\mathbf{K}_{\text{HH}}$, each of which is constructed as the outer product of the associated 1D filters: 

\begin{equation} \label{eq_anafilt}
\begin{aligned}
\mathbf{K}_{\text{LL}} = \ell \otimes \ell, \quad
\mathbf{K}_{\text{LH}} = \ell \otimes h, \\
\mathbf{K}_{\text{HL}} = h \otimes \ell, \quad
\mathbf{K}_{\text{HH}} = h \otimes h.
\end{aligned}
\end{equation}

Using these filters, the four ``subbands'' ($\mathbf{X}_\text{LL}$, $\mathbf{X}_\text{LH}$, $\mathbf{X}_\text{HL}$ and $\mathbf{X}_\text{HH}$) are obtained through convolution followed by downsampling, analogous to the 1D case. In deep learning frameworks, this operation can be implemented as a convolution with stride 2 \cite{MLWNet}. The LL subband captures the coarse structural content of the image. The LH and HL subbands emphasize horizontal and vertical details, respectively. The HH subband primarily represents diagonal details. 

The 2D Inverse DWT also follows analogously to the 1D case. The 2D synthesis filters are constructed from the 1D synthesis filters $\tilde{\ell}$ and $\tilde{h}$ via outer products, similar to Equation \eqref{eq_anafilt}. Starting from the subbands, the reconstructed image $\hat{\mathbf{X}}$ is also obtained through upsampling, followed by convolution with said filters.

The multi-level DWT is obtained by recursively applying the Forward DWT to the LL subband of the previous level, yielding a hierarchical multi-resolution representation. We visualize the DWT operations in Fig. \ref{fig_vis_dwt}.

\section{Methodology} \label{sec_method}

Figure~\ref{fig_framework} illustrates our multi-level DWT-based framework for frequency modulation. Firstly, the input image is decomposed by the Forward DWT using the 2-tap Haar wavelet. The resulting 1-level LL subband is decomposed again. Noticeably, we freeze all wavelet filters and introduce a learnable scaling parameter $\alpha$ that modulates the high-pass filters $h$. Since $\alpha$ is initialized to zero, all high-frequency (HF) subbands are zero-valued initially (blacked out in Figure~\ref{fig_framework}). The Inverse DWT is applied recursively to reconstruct the original image, which is highly coarse due to the absence of HF components. The resulting image is fed to 3DGS training and optimized with $\mathcal{L}_\text{3DGS}$, as introduced in Section \ref{sec_3dgs}. This represents the coarse stage of the modulation curriculum.

Another objective is required to optimize the wavelet towards PR, thus retrieving the full-spectrum image. For a 2-channel filter bank, the PR requirement first imposes the ``Alias Cancellation'' condition \cite{wavelet-book}:

\begin{equation}
\tilde{L}(z) H(-z) + \tilde{H}(z) L(-z) = 0
\end{equation}
where $L(z),H(z)$ and $\tilde{L}(z),\tilde{H}(z)$ denote the $z$-transforms of the 1D analysis and synthesis filters, respectively. Additionally, the ``No Distortion'' condition is required:

\begin{equation}
\tilde{L}(z) L(z) + \tilde{H}(z) H(z) = 2
\end{equation}

We convert these conditions into residual-based losses:
\begin{equation}
\mathcal{L}_{\text{alias}}
= \left\|
\tilde{L}(z)\, \alpha H(-z)
+ \tilde{H}(z)\, L(-z)
\right\|_2^2
\end{equation}
and
\begin{equation}
\mathcal{L}_{\text{dist}}
= \left\|
\tilde{L}(z)\, L(z)
+ \tilde{H}(z)\, \alpha H(z)
- 2
\right\|_2^2
\end{equation}
where only the scaling parameter $\alpha$, which modulates $H(z)$, is learnable and initialized to zero. The overall wavelet regularization objective is defined as:
\begin{equation}
\mathcal{L}_{\text{PR}}
=
\mathcal{L}_{\text{alias}}
+
\mathcal{L}_{\text{dist}}
\end{equation}
and the total training objective is as follows:
\begin{equation}
\mathcal{L}
=
\mathcal{L}_{\text{3DGS}}
+
\lambda_{\text{PR}}\mathcal{L}_{\text{PR}}
\end{equation}
where $\lambda_{\text{PR}}$ is the balancing term between the two objectives.

\section{Experiments}

\subsection{Dataset \& Implementation Details}

Our framework is evaluated on the LLFF dataset~\cite{LLFF} with 3 input views and on the Mip-NeRF 360 dataset~\cite{MipNeRF360} with 12 input views. These benchmarks consist of multiple multi-view images capturing various objects and scenes. The novel view synthesis capability is assessed using held-out test images.

Rendering quality is measured using PSNR, SSIM, and LPIPS~\cite{LPIPS}. To quantify Gaussian optimization, we record the peak number of Gaussians during densification stages and average this value across all scenes in each dataset. We also report the average training time per scene.

Comparisons are conducted against Vanilla 3DGS \cite{3DGS}, Opti3DGS \cite{Opti3DGS} and AutoOpti3DGS \cite{AutoOpti3DGS}, with all methods trained for 10K iterations under identical hyperparameters for fair comparison. We re-adopted AutoOpti3DGS but implemented the DWT via stride-2 convolutions following \cite{MLWNet}, enabling PyTorch's autograd to directly optimize the 1D filter parameters. The original implementation realizes the DWT via Toeplitz-structured matrix multiplication following \cite{WaveCNet}, requires hand-writing custom backpropagation equations, and optimizes the entire 2D filter. Under this re-implementation, AutoOpti3DGS becomes a special case of our framework that leverages 1-level DWT only, and that optimizes the full high-pass analysis filter $h$. For our extension, we only learn the scaling parameter $\alpha$, and set its learning rate to 1e-4. The balancing weight $\lambda_{\text{PR}}$ is 0.05.

\begin{figure}[htbp]
    \centering
    \includegraphics[width=\linewidth]{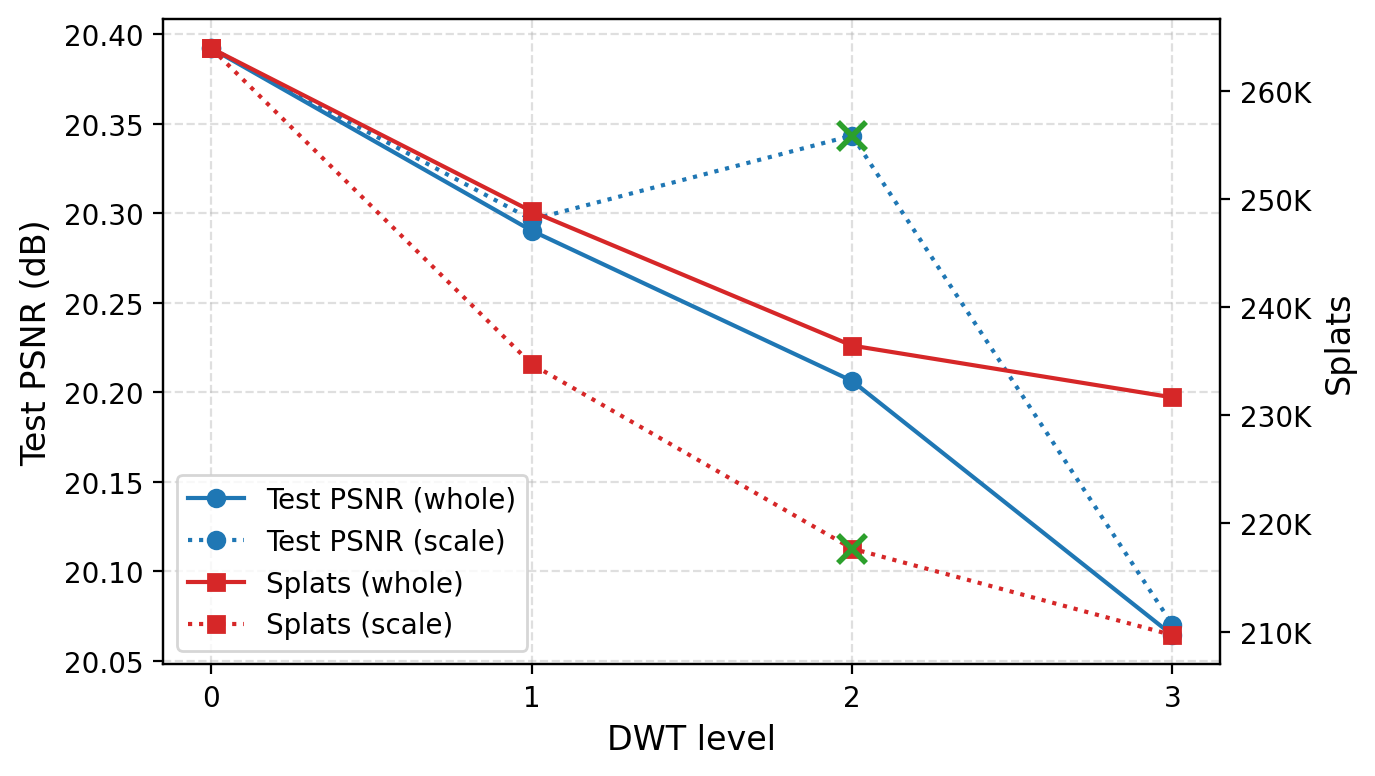}
    \caption{Ablation results on DWT levels and scaling parameter effects (3-view LLFF \cite{LLFF} dataset).}
    \label{fig_ablation}
\end{figure}

\subsection{Quantitative Results}

We provide the quantitative results at Tables \ref{tab_results_auto3dgs_llff} and \ref{tab_results_auto3dgs_mipnerf360}. Generally, compared to the 3DGS baseline, our multi-level strategy proves to reduce Gaussian counts significantly, by $\sim$50K and $\sim$100K Gaussians for the LLFF and Mip-NeRF 360 dataset, respectively. Compared to the closest efficient baseline, 1-level DWT-based AutoOpti3DGS, our method further reduces $\sim$30K Gaussians due to the expanded frequency modulation curriculum. Our method also incurs some additional training time, $\sim$10-20 seconds, because the multi-level DWT requires more computation than the 1-level version. We plan to tackle this in future works using lazy regularization, where $\mathcal{L}_{\text{PR}}$ is evaluated only every few iterations.

\begin{table}
\centering
\renewcommand\tabularxcolumn[1]{>{\RaggedRight\arraybackslash}p{#1}}
\begin{tabularx}{\linewidth}{lccccc}
\toprule
&\multicolumn{1}{X}{PSNR ($\uparrow$)}
&\multicolumn{1}{X}{SSIM ($\uparrow$)}
&\multicolumn{1}{X}{LPIPS ($\downarrow$)}
&\multicolumn{1}{X}{\#G ($\downarrow$)}
&\multicolumn{1}{X}{Time ($\downarrow$) (s)}
\\
\toprule
Opti3DGS \cite{Opti3DGS} & 19.59 & 0.660 & 0.228 & 247K & \textbf{105} \\
AutoOpti3DGS \cite{AutoOpti3DGS} & 20.29 & 0.703 & 0.200 & 249K & 131 \\
Ours & 20.34 & 0.687 & 0.222 & \textbf{218K} & 142 \\
\midrule
3DGS \cite{3DGS} & \textbf{20.40} & \textbf{0.706} & \textbf{0.197} & 272K & 109 \\
\bottomrule
\end{tabularx}
\vspace{2mm} 
\caption{Quantitative results, 3-view LLFF \cite{LLFF} dataset}
\label{tab_results_auto3dgs_llff}
\end{table}

\begin{table}
\centering
\renewcommand\tabularxcolumn[1]{>{\RaggedRight\arraybackslash}p{#1}}
\begin{tabularx}{\linewidth}{lccccc}
\toprule
&\multicolumn{1}{X}{PSNR ($\uparrow$)}
&\multicolumn{1}{X}{SSIM ($\uparrow$)}
&\multicolumn{1}{X}{LPIPS ($\downarrow$)}
&\multicolumn{1}{X}{\#G ($\downarrow$)}
&\multicolumn{1}{X}{Time ($\downarrow$) (s)}
\\
\toprule
Opti3DGS \cite{Opti3DGS} & 19.19 & 0.552 & 0.360 & 636K & \textbf{151} \\
AutoOpti3DGS \cite{AutoOpti3DGS} & 19.24 & 0.541 & 0.381 & 615K & 182 \\
Ours & 19.29 & 0.560 & 0.355 & \textbf{589K} & 200 \\
\midrule
3DGS \cite{3DGS} & \textbf{19.30} & \textbf{0.564} & \textbf{0.352} & 701K & 155 \\
\bottomrule
\end{tabularx}
\vspace{2mm} 
\caption{Quantitative results, 12-view Mip-NeRF 360 \cite{MipNeRF360} dataset}
\label{tab_results_auto3dgs_mipnerf360}
\end{table}

\subsection{Ablation Study}

Figure \ref{fig_ablation} presents the ablation results. The solid lines represent learning both filter coefficients, similar to AutoOpti3DGS (``whole'' mode in the legend). The dotted ones represent learning the scaling parameter only (``scale'' mode). The latter consistently result in lower Gaussian counts, and potentially better PSNR, as seen in the 2-level. We hypothesize this is because learning only the scaling parameter causes less gradient conflicts, so 3DGS can focus on the reconstruction task. Furthermore, increasing DWT levels leads to progressively decreasing Gaussian counts, at the expense of PSNR drops. We hypothesize that deeper levels produce overly coarse initial reconstructions with block-like artifacts that hamper reconstruction, especially as the Haar is used. There is a compromise between DWT levels and rendering quality. The best-performing configuration is marked with the green ``X'' symbols.

\section{Conclusion}

We propose a learnable multi-level DWT-based frequency modulation framework for 3DGS that consistently reduces Gaussian counts while preserving rendering quality. The modulation can be performed using only a single scaling parameter, rather than learning the full 2-tap high-pass filter. Experiments demonstrate notable Gaussian reductions over Vanilla 3DGS and 1-level DWT baselines with competitive rendering performance.




\bibliographystyle{IEEEtran}
\bibliography{main}

\end{document}